\documentclass[manuscript]{emulateapj} 
\usepackage[normalem]{ulem}
\usepackage{color}

\def \rozo{{\it Rozo}}
\def \johnston{{\it Johnston}}
\def\hmsun{\; h^{-1} \; M_{\odot}}
\def\msun{\;M_{\odot}}
\def\mass{M}

\newcommand{\myemail}{tomaszbi@umich.edu}
\newcommand{\avg}[1]{\left\langle #1 \right\rangle}

\renewcommand{\thefootnote}{\fnsymbol{footnote}}

\slugcomment{To Appear In ApJ}

\shorttitle{Impact of Systematics on SZ-Optical Scaling Relation}
\shortauthors{Biesiadzinski et al.}

\begin{document}

\title{Impact of Systematics on SZ-Optical Scaling Relations}

\author{T. Biesiadzinski$^{1}$\footnote{E-mail: \myemail}, J. McMahon$^{1}$, 
C. Miller$^{2}$, B. Nord$^{1}$\footnote{AGEP Fellow} and L. Shaw$^{3}$}

\affil{$^{1}$Physics Department, University of Michigan, Ann Arbor, MI 48109
\\$^{2}$Astronomy Department, University of Michigan, Ann Arbor, MI 48109
\\$^{3}$Physics Department,Yale, New Haven, CT 06511}

\begin{abstract}
One of the central goals of multi-wavelength galaxy cluster cosmology is to unite all cluster observables to form a consistent understanding of cluster mass. Here, we study the impact of systematic effects from optical cluster catalogs on stacked SZ signals. We show that the optically predicted $Y$-decrement can vary by as much as 50\% based on the current $2\sigma$ systematic uncertainties in the observed mass-richness relationship. Mis-centering and impurities will suppress the SZ signal compared to expectations for a clean and perfectly centered optical sample, but to a lesser degree. We show that the level of these variations and suppression is dependent on the amount of systematics in the optical cluster catalogs. We also study X-ray luminosity-dependent sub-sampling of the optical catalog and find that it creates Malmquist bias increasing the observed $Y$-decrement of the stacked signal. We show that the current Planck measurements of the $Y$-decrement around SDSS optical clusters and their X-ray counterparts are consistent with expectations after accounting for the $1 \sigma$ optical systematic uncertainties using the \johnston\  mass richness relation.
\end{abstract}


\section{Introduction}

Measurements of the abundance of galaxy clusters as a function of their masses and redshift provides an important constraint on the nature of dark matter and dark energy \citep[e.g.,][]{vik09,v10,seh11}. Joint analysis of multi-wavelength observations, including optical cluster catalogs and their Sunyaev-Zel'dovich \citep[SZ--][]{sz72,birk99,carl02} counterparts, will help realize the full cosmological potential of galaxy clusters \citep{cunha09,rozo09,wu10}.

Optical galaxy cluster surveys have identified thousands of clusters down to a mass limit of $\sim 10^{14} \msun$ \citep[e.g., the maxBCG catalog of SDSS clusters,][]{koest07a}, and millimeter wave surveys have discovered hundreds of clusters using the SZ effect \citep[e.g.,][]{v10,will11,act11,ade11a}, albeit to a higher mass limit due to instrumental noise. Using these catalogs, researchers apply \emph{mass-observable relations} to relate the true underlying halo mass to observed properties, like the galaxy member count in the optical (richness, $N_{gal}$ or $N_{200}$) or the SZ decrement ($y$ or $Y_{\rm 500}$).  

One can probe scaling relations down to masses below the detection limit by stacking the signal around known clusters. For instance, stacking has been used to the great benefit of mass calibration in weak lensing and X-ray studies \citep[]{shel07,rykoff08}. Similarly, the SZ/X-ray cluster scaling-laws and pressure-profiles were evaluated by Komatsu et al. (2011) and Melin et al. (2011), who stacked the SZ signal from WMAP data around known optical/X-ray clusters. These joint optical/X-ray/SZ analyses allow researchers to take advantage of the large volumes and mass ranges from optical cluster catalogs in combination with the lower scatter in the mass observable relation in X-ray/SZ catalogs \citep{shaw08, nagai06, rasia11, motl05}.
 
The SZ signal recovered from stacking Planck data at positions of the maxBCG \citep{koest07a} clusters shows a deficit of SZ signal compared to what is expected from current mass-richness scaling relationships \citep{planck11}; this discrepancy has been confirmed using WMAP data \citep{draper11}. This discrepancy manifests itself differently for two mass-richness calibrations \citep{john07,rozo09} both of which are based on the \cite{shel07} stacked weak-lensing mass measurements of the maxBCG clusters. For the \cite{john07} calibration, a simple reduction in the global weak-lensing mass calibration by 25\% would eliminate the discrepancy.  The \cite{rozo09} mass calibration requires a larger correction and a scaling law that is not self-similar.  \cite{planck11} also show that a subset of the maxBCG clusters with measured X-ray luminosities from the MCXC catalog \citep{piff10} can match the predicted $Y_{\rm 500}$ vs. richness scaling relationship, although they did not consider selection effects inherent in such a hybrid catalog.

The \cite{planck11} analysis was based on a comparison of the observed $Y_{\rm 500}$ around maxBCG clusters to two models with different mass-richness calibrations and without including optical systematics. They evaluated the impact of impurities in the optical catalog as well as scatter in the mass-richness relations and concluded that neither could account for the observed discrepancy individually. Here, we broaden the \cite{planck11} analysis to include the {\it  uncertainties in the mass calibrations} as well as the {\it combined systematic effects} in optical cluster catalogs. Instead of two model predictions to compare against the data, we look at the family of predictions which come from uncertainties in the calibrations and the ranges of systematics in optical cluster catalogs.

There are numerous systematic effects in optical galaxy cluster catalogs.  These include the cluster selection (as a function of mass $\mass$ and redshift $z$) which comprises: \emph{completeness}--- the probability that a true halo will be detected; and \emph{purity}--- the probability that a detection correctly identifies a halo rather than noise \citep[e.g.,][]{miller05}.   Cluster \emph{redshifts} estimated using photometric data are uncertain which introduces scatter in the observed redshift.  There is  uncertainty in the \emph{mass-richness} calibration as well as \emph{scatter}.  Finally, mis-identification of BCGs in the maxBCG cluster-finder produces angular offsets between true and recovered \emph{cluster centers} \citep{john07} called \emph{mis-centering}. Centering offsets driven by other mechanisms \citep[e.g., astrophysical:][]{sand09} are smaller than those caused by the BCG mis-identification, and so we do not consider them in this study.

Using mock clusters taken from N-body simulations, we directly manipulate the purity, mass-scatter, scaling calibrations and their uncertainties. We then re-create the Planck richness stacking technique on these mock catalogs to create model $Y_{\rm 500}$-richness relations and compare to the Planck observations. In \S\ref{sec:simulations}, we describe the N-body simulations and the suite of simulated optical cluster catalogs with various systematics, the mock Planck SZ observations, the mock X-ray observations and the stacking procedure.  We then show the results of stacking the SZ signal for each systematic to explore how each systematic can individually affect the SZ signal (\S\ref{sec:res}), and we compare to the Planck joint SZ-optical and X-ray analyses (\S\ref{sec:simjoint}). Throughout this paper, we assume a $\Lambda$CDM cosmology with a $\Omega_{\Lambda}$=0.75 and $H_{0}$=0.71 unless otherwise noted.

\section{Simulations}
\label{sec:simulations}
We begin with a simulated mass function and halo positions from an N-body lightcone.  We then impose observables and realistic systematic effects to produce mock optical catalogs and then dress the halos with gas and simulate Planck SZ observations.
\subsection{N-body Lightcone}
\label{nbodysection}
To generate the mock SZ maps and galaxy catalogs, we begin with the halo positions from a large (${\rm N} = 1260^3$ particles, 1000 [Mpc $h^{-1}$]$^3$) cosmological dark matter simulation. Cosmological parameters were chosen to be consistent with those measured from the five-year {\it WMAP} data \citep{dunkley09} combined with large-scale structure observations, namely $\sigma_8 = 0.8$, $\Omega_M = 0.264$ and $\Omega_b = 0.044$. The simulation was carried out using the tree-particle-mesh code of \citet{bode03}. In total, the lightcone covers a single octant on the sky ($\sim 5000$ deg$^2$) to a redshift of 3, containing halos with masses $\mass_{\rm  FOF} > 3 \times 10^{13} \hmsun$.  

The simulation does not provide any observables (e.g., richness or SZ/X-ray luminosity). We do not use the halo masses output from this particle simulation directly but rather use the procedure described in Section \ref{halosimsection}. With the mass resolution available from this simulation we can reproduce the properties of the maxBCG catalog, including systematics, for clusters with $M_{500} > 6 \times 10^{13}\ \hmsun$ or $N_{200} > 20$.

\subsection{Simulated Halo Catalogs}
\label{halosimsection}

In the Planck analysis \citep{planck11}, the clusters are binned according to their optical richness ($N_{200},\ N_{Gals}$). Richness is defined as the number of bright red galaxies (within the E/S0 ridgeline) inside $R_{200}$ that are brighter than 0.4 L$^*$ \citep{koest07a}. Recall, that richness is an observed quantity and at any fixed value clusters can have a range of true masses (the {\it mass scatter}).

The halo catalog provides a mass function and large-scale structure according to our chosen cosmology and similar to the observed universe. We cannot directly assign richnesses to these halo masses that match the observed scatter. Therefore we create a mock catalog of masses and richnesses and assign them to the N-body halos to preserve the large scale structure of the universe.

For each richness we center a Gaussian probability distribution function (PDF) in $ln(Mass)$. The center of the Gaussian is taken from a particular scaling relation; the width represents the scatter in $ln(Mass)$ at fixed richness. We draw from these Gaussian PDFs to create a list of masses including scatter for each richness. We adjust the number of draws from each PDF to reproduce the halo mass function (e.g., we draw more times from the low mass bins). This provides a table of richnesses and associated masses with the same halo mass function as the N-body simulation. We sort the N-body halos and this table by mass. We associate the positions of the N-body halos to the drawn table based on this ordering. This produces our mock cluster catalog which includes large-scale structure and reproduces a particular choice of scaling relations and scatter. We use these masses to create the SZ profiles (Section \ref{sec:mocksz}) and X-ray luminosities (Section \ref{sec:mcxcsim}).

\subsection{Optical Cluster Catalogs}
\label{opticalsimsection}

Using the procedure described above we create mock catalogs that are modified as follows to include systematic effects:
\begin{enumerate}

\item {\bf Mass-richness Calibration:} We varied the richnesses of the halos according to equation 26 (and associated uncertainties) from \cite{john07}:
\begin{equation}
\avg{M_{200}|N_{200}} = M_{200|20} \left (\frac{N_{200}}{20}\right)^{\alpha_{N}} \\
\end{equation}
\begin{flushright}
$M_{200|20} = (8.8 \pm 0.4_{stat} \pm 1.1_{sys}) \times 10^{13} \hmsun$
$\alpha_{N} = 1.28 \pm 0.04 $
\end{flushright}
or equation 4 from \cite{rozo09}:
\begin{equation}
\frac{\avg{M_{500}|N_{200}}}{0.71 \times 10^{14}\ \hmsun} = \exp(B_{M|N}) \left (\frac{N_{200}}{40}\right)^{\alpha_{M|N}} \\
\end{equation}
\begin{flushright}
$\alpha_{M|N} = 1.06 \pm 0.08_{stat} \pm 0.08_{sys}$
$B_{M|N} = 0.95 \pm 0.07_{stat} \pm 0.10_{sys}$
\end{flushright}

We look at one and two $\sigma$ deviations from these mass calibrations. Masses are converted from $M_{200}$ to $M_{500}$ assuming an NFW profile and mass concentrations from \cite{duffy08}, and are relative to the critical density. 

\item {\bf Completeness:} We vary the fraction of halos in bins of redshift and mass. 

\item {\bf Purity:} We add into the halo catalogs an additional number of false halos in bins of mass and redshift to create samples with different purities. We either vary the purity as a constant with mass and redshift or match the published maxBCG purity of \cite{koest07a}.

\item {\bf Redshifts:} We scatter the true halo redshifts by normal distributions with varying widths as large as $\sigma_z=0.05$.

\item {\bf Center Offsets:}  We offset the center for a fraction of the clusters according to Eq. 10 in  \cite{john07}. For the offset clusters, the actual amount of the offset is described by Eq 8 in  \cite{john07}. See also Figures 4 and 5 in \cite{john07}.

\item {\bf Mass Scatter} We vary the width of the log-normal distribution of masses at fixed richness.

\end{enumerate}

Realizations of mock optical cluster catalogs are created to investigate the impact of individual systematic effects. These include maxBCG-like systematics \citep{koest07a,koest07b,john07,rozo09} and more general systematics that are constant in redshift and mass (or richness). We also create catalogs combining maxBCG systematics to compare to data. In our maxBCG-like mocks, the fraction of incorrectly centered clusters ranges from 12\% in the highest richness bins to 39\% in the lowest richness bins with a mean offset of 0.6Mpc corresponding to $3'$ for a cluster at the mean redshift of z = 0.2; the completeness and purity are $>$90\% above M$_{500}$ $>\ 1 \times 10^{14}\ \hmsun$ and have an estimated uncertainty of 2.5\%; the mass scatter is 0.45 $\pm$ 0.10, similar to \rozo\ ($\sigma_{\ln(M)|N_{200}}=0.45^{+0.20}_{-0.18}$ ($95\%$ CL) at $N_{200}\approx 40$).

\subsection{Mock SZ Sky Maps and the Stacked Signal}
\label{sec:mocksz}
The halo SZ signals are generated using a thermal pressure profile suggested by \citep{arnaud10} and used in the Planck maxBCG stacking analysis \citep{planck11}. \cite{bon11} compares the pressure profiles of 25 massive relaxed clusters observed in X-ray and with the Sunyaev-Zel'dovich Array (SZA) and find that they agree well with the \cite{arnaud10} profile up to $R_{500}$. We project the profile along the line-of-sight to produce a compton-Y profile, scaled to the appropriate size for each halo redshift. Mock {\it Planck} observations were created in each frequency band using the appropriate beam sizes, instrument noise and primary CMB \citep{ade11a} temperature anisotropy. We concluded that the 143 GHz channel reproduced the dominant features of the multi-frequency analysis, and so we restricted our analysis solely to this channel, which has a beam size of 7.18 arcminute FWHM and a noise of 0.9 $\mu K-$degree. 

At the position of each optical cluster, we extracted the integrated thermal SZ signal $Y_{500}$ from each SZ sky map using a matched filter \citep[e.g.,][]{her02,mel06} with an Arnaud profile \citep{arnaud10}, the size of which is inferred from either the \johnston\ or \rozo\ richness-mass scaling relations \citep[same as used in][]{planck11}. We stacked these match filtered signals in richness bins; then the amplitude is calibrated by comparing the spherical $Y_{500}$ of the halos with the amplitude in the stacked SZ signal in the absence of systematics. We found that including an intrinsic scatter of 25\% in $Y_{500}-M_{500}$ \citep{shaw08} did not affect our results beyond increasing statistical uncertainties in individual catalog realizations and so we did not include this additional scatter in the following analysis.
\subsection{maxBCG-MCXC Subsample}
\label{sec:mcxcsim}
The Planck team studied a subset of the maxBCG catalog whose positions were matched to within $\sim$3 arcminutes of X-ray clusters from the MCXC catalog \citep{piff10}. Starting with the masses of our simulated halos we assign X-ray luminosities ($L_{X}$) and scatter according to Table 1 in \cite{arnaud10}. We reproduce the scatters in the $L_{X}$ scaling relations (at fixed mass and richness) that are observed in \cite{rozo09} where $\sigma_{\ln(L)|M}$ ranges from $0.5$ at low mass to $0.45$ at high mass and $\sigma_{\ln(L)|N}$ is a constant $0.85$ at all richnesses. We also vary the input $L_{X}$ and scatter to asses the sensitivity to those parameters. We then select subsets of the simulated halos which have the same redshift and $L_{X}$ distribution as the MCXC subsample. This allows us to reproduce the MCXC subsample without needing to characterize the exact selection function which is undoubtedly complex as this catalog is drawn from heterogeneous X-ray data.  We also ensure that the mis-centering for this mock MCXC-maxBCG catalog is truncated at  3'. The maxBCG-MCXC mock catalogs need not have the same scatter in the mass-richness relation as we imprinted into the full maxBCG mock samples. This is because we imprint the observed scatter from Arnaud et al. (2010) directly onto the full catalog and then draw a sub-sample. For the MCXC/maxBCG mock subsamples, $\sigma_{\ln(L)|N}$ drops to $0.70$ and  $\sigma_{\ln (M)|N_{200}}$ drops to $0.40$. 

\subsection{Correlations of Observables}
\label{sec:correlations}

\begin{figure}
\begin{center}
\includegraphics[width=80mm,height=70mm]{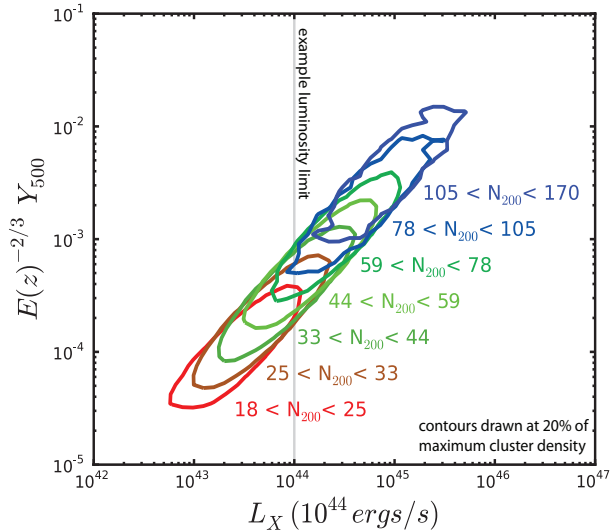}
\caption{Correlation between simulated X-ray luminosities ($L_{X}$) and SZ signals ($Y_{500}$) in the richness bins used in this work. The contours are drawn where the number of clusters is 20\% as large as the number at the mean value of $L_{X}$ and $Y_{500}$ for each richness bin (the center of each contour). The gray vertical line illustrates the approximate limit in X-ray luminosity reached by some of the surveys used in the construction of the MCXC \citep{piff10} where they overlap with maxBCG.}
\label{fig:lx_ysz_corr}
\end{center}
\end{figure}

The large scatter in true mass at fixed richness (see Section \ref{opticalsimsection}) induces a correlation between the observed X-ray luminosity and SZ signal. We note that this correlation is distinct from a secondary correlation in the scatter of observables. Figure \ref{fig:lx_ysz_corr} shows the cluster density as a function of $L_{X}$ and $Y_{500}$ in various richness bins. Within each richness bin there is a strong correlation between the two observables. This will be crucial for understanding the joint maxBCG-MCXC subsample discussed in Section \ref{xraysample}.

Our simulation pipeline does not create correlated scatters in the observables at fixed mass. Such correlations are expected due to common substructue within clusters and projection effects \citep{white10} however they are likely secondary effects and are beyond the scope of this work (see \cite{ang12}).

\centerline{}

\section{Results} \label{sec:res}

\begin{figure}
\begin{center}
\includegraphics[width=85mm,height=106mm]{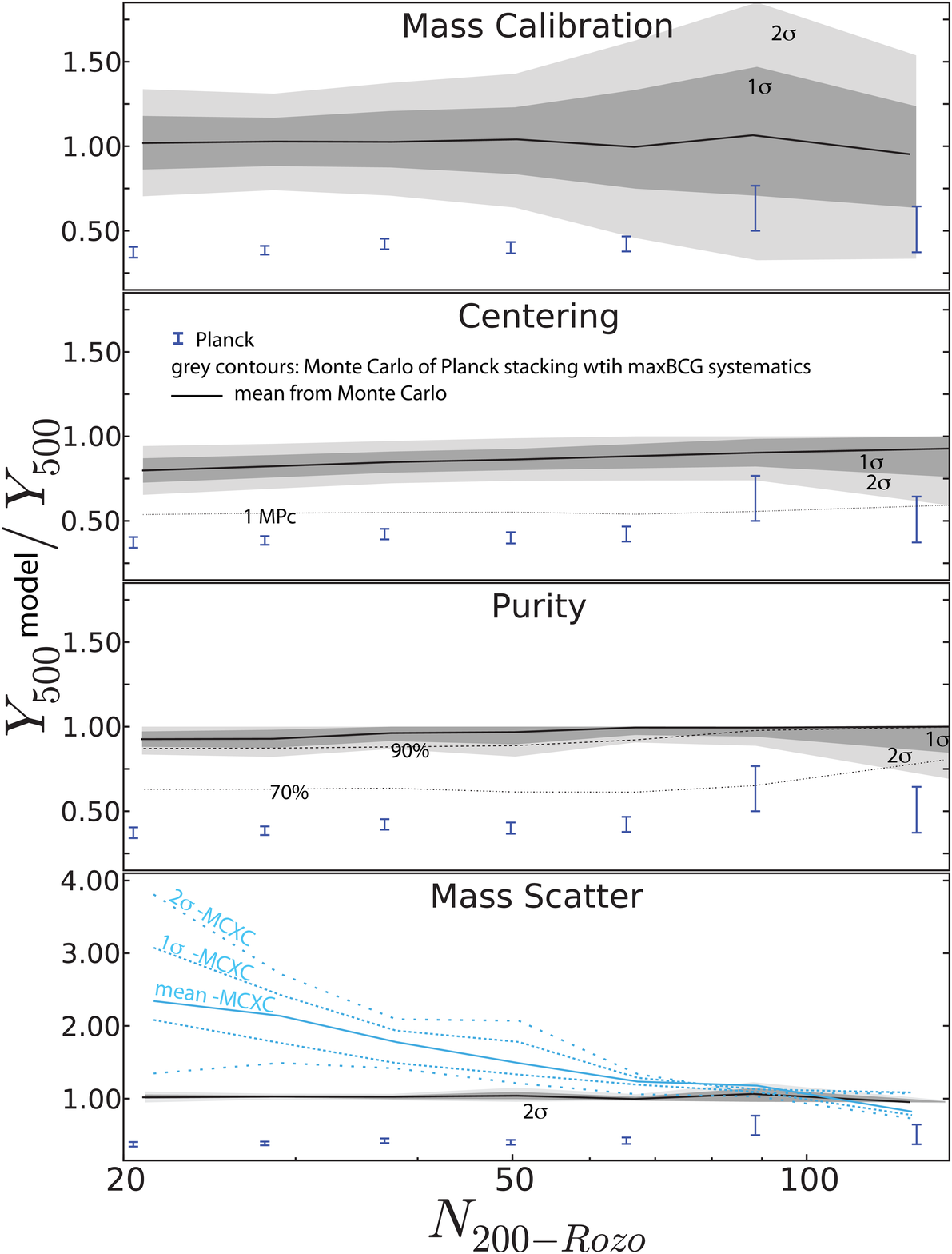}
\caption{A comparison of the stacked $Y_{500}$ in our family of mock cluster catalogs to a single ``perfect'' cluster catalog that has been calibrated according to \cite{rozo09}. The solid black lines show the model with maxBCG-like systematics included (individually). The gray bands show the range of models after we include the 1 and 2 $\sigma$ uncertainties on the individual optical systematics in addition to statistical uncertainties. Gray dotted lines show more general models, while the blue lines in the bottom panel are specific to the maxBCG/MCXC sub-sample. The error bars are the Planck data. Uncertainty in the mass calibration is the dominant effect on the model predictions, however impurity and mis-centering both bias the model predictions towards lower values of $Y_{500}$. On the other hand, X-ray luminosity selected sub-samples (e.g., the MCXC) show highly biased $Y_{500}$ predicted values (compared to a perfect optical catalog). See Figures \ref{fig:maxbcg_and_planck} and \ref{fig:maxbcg_and_mcxc} for the combined effects of these systematics.}
\label{fig:systematics}
\end{center}
\end{figure}
In Figure \ref{fig:systematics}, we compare the stacked $Y_{500}$ in our family of mock cluster catalogs to a ``perfect'' cluster catalog that has been calibrated according to \cite{rozo09}. The ``perfect'' catalog uses a single calibration and does not contain any of the systematics we discuss in Section \ref{opticalsimsection}. This is identical to the model the Planck team used to compare to the data \citep{planck11}. In each panel, the solid black line shows the average ratio (over multiple mock realizations) for models which apply the fiducial maxBCG values for calibration, mis-centering, purity, and mass scatter individually (as described at the end of Section \ref{opticalsimsection}). The gray bands show the range of models using the 1 and 2 $\sigma$ uncertainties on those parameters. Dotted-lines show more general models (e.g., 70\% purity independent of mass). We also show the Planck data presented in \cite{planck11}. We do not show redshift scatter and completeness since we found them to have negligible effects at maxBCG levels. 

Systematic uncertainties (2$\sigma$) in the mass-richness calibration result in up to 50\% range in the model $Y_{500}$ measurements. This is because the $Y_{500}$ values from our perfect catalog are calculated from a single mass calibration, while the model $Y_{500}$s are calculated using the masses drawn from the calibration including 1 and 2 $\sigma$ uncertainties.

Mis-centering suppresses (biases low) the model $Y_{500}$s over the entire mass range, with the largest effect at low mass ($\sim$ 25\% suppression). This can be understood from the convolution of the Planck beam ($\sim$ 7' full width at half of maximum) and the centering offsets which are on average $\sim$ 3' at the median redshift of the optical sample. The offsets are large compared to the Planck beam, which blurs out the SZ-signal after the convolution. The impact of this effect increases to $\sim 25\%$ at low mass, since the maxBCG mis-centering fraction is mass dependent.

Impurities suppress (biases low) the amplitude of the model $Y_{500}$s by introducing pure noise into the SZ maps. As also noted by \cite{planck11}, high levels of impurity would be required to explain the discrepancy with the data. Just as important, the weak-lensing calibration of the mass-richness relation would also be affected by large impurities which would lead to an enhancement in the mass-richness relation. Since $Y_{500}\sim \mass^{\frac{5}{3}}$, high impurities could even cause the observed SZ signal to be enhanced compared to the systematics-free case (something neither we nor Planck detect). Accurate modeling of the impact of impurities on $Y_{500}$ requires simulating its effect on the weak-lensing calibration of the optical catalog.

The stated uncertainty in mass scatter \citep{rozo09} does not have a significant impact on the SZ signal recovered using a maxBCG-like catalog (see the gray band in the bottom panel of Figure \ref{fig:systematics}). However, the same can not be said for the MCXC-like subsample (blue lines in the same panel and see Section \ref{sec:mcxcsim}). The X-ray selection causes a Malmquist bias in low richness bins where the X-ray sub-sample preferentially contains brighter (and thus  more massive) clusters. Figure \ref{fig:lx_ysz_corr} illustrates that selecting clusters above some $L_{X}$ limit (like the example shown by the gray line) preferentially selects clusters with high $Y_{500}$. Larger mass scatter increases the correlation between $L_{X}$ and $Y_{500}$ and therefore enhances the Malmquist bias.
Richness bins that lie completely to the right of the $L_{X}$ limit are not affected by this bias and so the SZ signal there is not enhanced. 

\subsection{Simulating Planck-maxBCG Joint Analysis}\label{sec:simjoint}

\begin{figure}
\begin{center}
\includegraphics[width=85mm,height=47mm]{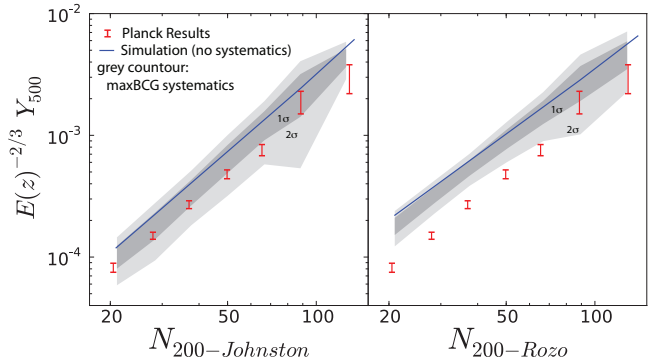}
\caption{The Planck data (error bars) compared to the single perfect model used in \citep{planck11} (blue line) and to the range of models (gray bands) after jointly combining all of the individual systematic effects seen in Figure \ref{fig:systematics}. The naive perfect model predicts higher (on average) $Y_{500}$ values compared to the models which include catalog systematics. The data are consistent with our model predictions within 1 $\sigma$ for the \johnston\ mass calibration.}
\label{fig:maxbcg_and_planck}
\end{center}
\end{figure}

Figure \ref{fig:maxbcg_and_planck} compares the Planck results to our models. The Planck data (error bars) are the same in both panels from \cite{planck11}. The solid blue lines shows the single naive perfect model based on either the \johnston\ (left) or \rozo\ (right) mass calibration in the absence of systematics \citep{planck11}. The gray bands show model predictions based on our Monte-Carlo mock cluster catalog realizations which include all of the maxBCG optical catalog properties, uncertainties, and systematics shown in Figure \ref{fig:systematics} and which were applied in the original weak-lensing richness mass calibrations. While the Planck data are statistically inconsistent with the naive perfect model prediction, they lie at the lower edge of the models which include the $\sim$1 $\sigma$ systematic uncertainties for the \johnston\ mass calibration. 

\subsection{Simulating maxBCG-MCXC Joint Sample}
\label{xraysample}

\begin{figure}
\begin{center}
\includegraphics[width=85mm,height=47mm]{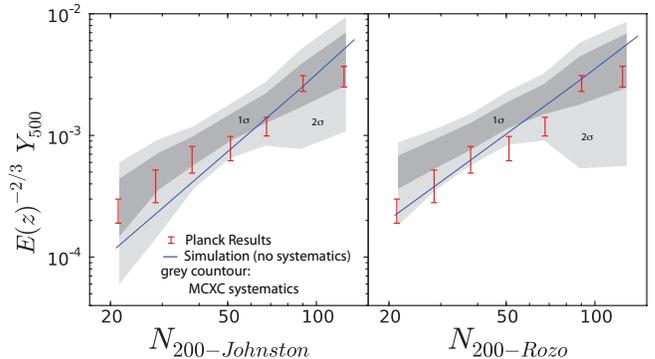}
\caption{The Planck data for the maxBCG/MCXC X-ray sub-sample (error bars) compared to the single perfect model used in \citep{planck11} (blue line) and to the range of models (gray bands) after jointly combining all of the individual systematic effects seen in Figure \ref{fig:systematics}. While the perfect model is the same as in Figure \ref{fig:maxbcg_and_planck}, the gray bands here include the bias seen in Figure \ref{fig:systematics} (bottom), which is caused after sub-sampling clusters based on their X-ray luminosities to match the observed data. The naive perfect model predicts lower (on average) $Y_{500}$ values compared to the models which include catalog systematics. The data are consistent with our model predictions at the 1 (2) $\sigma$ levels on the optical systematics for the \johnston\ (\rozo) mass calibration.}
\label{fig:maxbcg_and_mcxc}
\end{center}
\end{figure}

Figure \ref{fig:maxbcg_and_mcxc} shows our prediction for the MCXC sub-sample of the maxBCG catalog compared to the Planck data. The gray bands here include simulated optical and X-Ray systematics as well as the X-ray selection function. As expected from Figure \ref{fig:systematics}-bottom, we see a bias in the predicted $Y_{500}$ with decreasing richness due to the Malmquist bias present in low richness bins after only the brightest $L_{X}$ are selected (see Figure \ref{fig:lx_ysz_corr}). The Planck observations lie inside the lower edge of the models which include the 1 and $2 \sigma$ systematic uncertainties for \johnston\ and \rozo\ mass calibrations respectively.

\centerline{}

\section{Discussion}
The Planck team reported that the stacked SZ signal around optical clusters lies well below the single model expectation which does not include the optical catalog systematic uncertainties. On the other-hand, they find that the observed stacked $Y_{500}$ values around an X-ray limited sub-sample are consistent with the naive optical model. They concluded that the gas properties of clusters appear to be more stably related to each other than the gas-to-optical properties of clusters \citep{planck11}. In this work, we reach a fundamentally different conclusion: the $Y_{500}$ values observed by Planck are consistent with the model predictions for both the entire cluster sample and the X-ray sub-sample to within the 1 $\sigma$ optical systematic uncertainties of the \cite{john07} mass calibration. Not only do we argue that there is no significant discrepancy between the models and the observed Planck stacked $Y_{500}$ values around optical clusters, but we also argue that the optical and X-ray selected sub-samples simultaneously agree with model predictions. For instance, we can apply a single mass-richness calibration to the data and fit the predicted $Y_{500}$ models in Figures \ref{fig:maxbcg_and_planck} and \ref{fig:maxbcg_and_mcxc} simultaneously. However we do not pursue a joint SZ-optical mass calibration here, as it beyond the scope of this work.

We find that the dominant source of optical systematic uncertainty comes from the mass calibration, which alone can account for most of the original discrepancy noted by \cite{planck11}. Impurities and centering errors combine to bias the model predictions towards lower $Y_{500}$ for the optical samples while mass scatter biases the predictions high for low richness systems in the X-ray limited subsample. When fully accounted for, these systematics allow for models which are matched by the observed data for both the optical and X-ray cluster sub-samples in the Planck data. The range on the acceptable models is quite large and we note that the  SZ-optical scaling laws cannot by precisely characterized using this type of stacking until the optical systematics improve (specifically mass calibration and its scatter). 

This work highlights the importance of multi-wavelength studies of cluster properties as a source of cross-checks and a calibration. It is clear that optical systematics cannot be ignored and future analysis of stacked clusters should be done using Monte Carlo analysis to include a larger suite of systematic errors.

\acknowledgments
Acknowledgments: This work was supported through  DoE Grant DE-FG02-95ER40899. In addition, we are grateful for the support of the NSF-funded Michigan AGEP Alliance program. L. Shaw acknowledges the support of Yale University and NSF
grant AST-1009811. We would also like to thank Eduardo Rozo for comments that helped improve the clarity of this work.




\begin{thebibliography}{}

\bibitem[Ade et al.(2011a)]{ade11a} Ade, P. A. R., et al. [Planck Collaboration]
2011a, \aap, 536, A8


\bibitem[Ade et al.(2011c)]{ade11c} Ade, P. A. R., et al. [Planck Collaboration]
2011c, \aap, 536, A11

\bibitem[Aghanim et al.(2011a)]{planck11} Aghanim, N. et al. [Planck Collaboration]
2011a, \aap, 536, A12


\bibitem[Angulo et al.(2012)]{ang12} Angulo, R. E., Springel, V., White, S. D. M., Jenkins, A., Baugh, C. M., \&  Frenk, C. S. 2012, arXiv:1203.3216

\bibitem[Arnaud et al.(2010)]{arnaud10} Arnaud, M., Pratt, G. W.,
Piffaretti, R., B\"{o}hringer, H., Croston, J. H., \& Pointecouteau, E. 2010,
\aap, 517, A92

\bibitem[Birkinshaw(1999)]{birk99} Birkinshaw, M. 1999, \physrep, 310, 97

\bibitem[Bode \& Ostriker(2003)]{bode03} Bode, P., \& Ostriker, J. P. 2003,
\apjs, 145, 1


\bibitem[Bonamente et al.(2011)]{bon11} Bonamente, M., et al. 2011, arXiv:1112.1599

\bibitem[Carlstrom et al.(2002)]{carl02} Carlstrom, J. E., Holder, G. P., 
\& Reese, E. D. 2002, \araa, 40, 643

\bibitem[Cunha et al.(2009)]{cunha09} Cunha, C., et al. 2009, Phys. Rev. D., 79, 63009

\bibitem[Draper et al.(2011)]{draper11} Draper, P., Dodelson, S., Hao, J.,
\& Rozo, E. 2011, arXiv:1106.2185 

\bibitem[Duffy et al.(2008)]{duffy08} Duffy, A. R., Schaye, J., Kay, S. T., 
\& Dalla Vecchia, C. 2008, \mnras, 390, L64

\bibitem[Dunkley et al.(2009)]{dunkley09} Dunkley, J., et al. 2009,
\apjs, 180, 306


\bibitem[Herranz et al.(2002)]{her02} Herranz, D., Sanz, J. L., Hobson, M. P.,
Barreiro, R. B.,
Diego, J. M., Martínez-González, E., \& Lasenby, A. N. 2002, \mnras, 336, 1057

\bibitem[Johnston et al.(2007)]{john07} Johnston, D. E., et al. 2007,
arXiv:0709.1159 
 
\bibitem[Koester et al.(2007a)]{koest07a} Koester, B. P.,
et al. 2007a, \apj, 660, 239

\bibitem[Koester et al.(2007b)]{koest07b} Koester, B. P.,
et al. 2007b, \apj, 660, 221

\bibitem[Komatsu et al.(2010)]{kom10} Komatsu, E., et al. 2010,
arXiv:1001.4538 


\bibitem[Marriage et al.(2011)]{act11} Marriage, T. A.,
et al. 2011, \apj, 737, 61

\bibitem[Melin et al.(2006)]{mel06} Melin, J., Bartlett, J. G., \&
Delabrouille, J. 2006, \aap, 459, 341 

\bibitem[Melin et al.(2011)]{mel11} Melin, J., Bartlett, J. G.,
Delabrouille, J., Arnaud, M., Piffaretti, R., \& Pratt, G. W.
2011, \aap, 525, A139

\bibitem[Miller et al.(2005)]{miller05} Miller, C. J., et al. 2005, \aj, 130, 968

\bibitem[Motl et al. (2005)]{motl05} Motl, P.M. Hallman, E.J. and Burns, J.O. and Norman, M.L., 2005, \apjl, 623, L63

\bibitem[Nagai et al.(2006)]{nagai06} Nagai, D. 2006, \apjl, 650, 538

\bibitem[Piffaretti et al.(2011)]{piff10} Piffaretti, R., Arnaud, M.,
Pratt, G. W., Pointecouteau, E., \& Melin, J. 2011, \aap, 534, A109 

\bibitem[Rasia et al.(2011)]{rasia11} Rasia, E., Mazzotta, P., Evrard, A., Markevitch, M.,
       Dolag, K. and Meneghetti, M. 2011, \apjl, 729, 45

\bibitem[Rozo et al.(2009)]{rozo09} Rozo, E., et al. 2009, \apj, 699, 768

\bibitem[Rykoff et al.(2008)]{rykoff08} Rykoff, E.S, et al. 2008, Phys. Rev. D., 79, 63009

\bibitem[Sanderson et al.(2009)]{sand09} Sanderson, A. J. R., Edge, A. C.,
\& Smith, G. P. 2009, \mnras, 398,1698 

\bibitem[Sehgal et al.(2011)]{seh11} Sehgal, N.,
et al. 2011, \apj, 732, 44

\bibitem[Shaw et al.(2008)]{shaw08} Shaw, L. D., Holder G. P., \& Bode, P.
2008, \apj, 686, 206

\bibitem[Sheldon et al.(2007)]{shel07} Sheldon, E. S., et al. 
2009, \apj, 703, 2217


\bibitem[Sunyaev \& Zeldovich(1972)]{sz72} Sunyaev, R. A., \& Zeldovich, Y. B. 1972,
Comments on Astrophysics and Space Physics, 4, 173

\bibitem[Vanderlinde et al.(2010)]{v10} Vanderlinde, K.,
et al. 2010, \apj, 722, 1180

\bibitem[Vikhlinin et al.(2009)]{vik09} Vikhlinin, A.,
et al. 2009, \apj, 692, 1060

\bibitem[White et al.(2010)]{white10} White, M., Cohn, J. D., 
\& Smit, R. 2010, \mnras, 408,1818 


\bibitem[Williamson et al.(2011)]{will11} Williamson, R., et al. 
2011, arXiv:1101.1290

\bibitem[Wu et al.(2010)]{wu10} Wu, H.Y., et al. 2010, \apj, 713, 1207

\end{thebibliography}
\end{document}